\documentclass[sigconf]{acmart}
\usepackage[utf8]{inputenc}
\usepackage{adjustbox}
\usepackage{amssymb}
\usepackage{bbm}
\usepackage{bm}
\usepackage{enumitem}
\usepackage{booktabs} 
\usepackage{dsfont}

\usepackage{multirow}
\usepackage{array}
\newcolumntype{L}[1]{>{\raggedright\let\newline\\\arraybackslash\hspace{0pt}}m{#1}}
\newcolumntype{C}[1]{>{\centering\let\newline\\\arraybackslash\hspace{0pt}}m{#1}}
\newcolumntype{R}[1]{>{\raggedleft\let\newline\\\arraybackslash\hspace{0pt}}m{#1}}

\usepackage{colortbl}%

\newcommand{\myrowcolour}{\rowcolor[gray]{0.925}}
\setcopyright{rightsretained}

\copyrightyear{2017}
\acmYear{2017}
\setcopyright{acmcopyright}
\acmConference{KDD '17}{}{August 13-17, 2017, Halifax, NS, Canada}
\acmPrice{15.00}
\acmDOI{10.1145/3097983.3098061}
\acmISBN{978-1-4503-4887-4/17/08}

\begin{document}
\title{\textit{struc2vec}: Learning Node Representations from Structural Identity}

\author{Leonardo F. R. Ribeiro}
\affiliation{%
  \institution{Federal University of Rio de Janeiro}
  \institution{Systems Eng. and Comp. Science Dep.}
}
\email{leo@land.ufrj.br}

\author{Pedro H. P. Saverese}
\orcid{1234-5678-9012}
\affiliation{%
  \institution{Federal University of Rio de Janeiro}
  \institution{Systems Eng. and Comp. Science Dep.}
}
\email{savarese@land.ufrj.br}

\author{Daniel R. Figueiredo}
\affiliation{%
  \institution{Federal University of Rio de Janeiro}
  \institution{Systems Eng. and Comp. Science Dep.}
}
\email{daniel@land.ufrj.br}

\begin{abstract}
Structural identity is a concept of symmetry in which network nodes are identified according to the network structure and their relationship to other nodes. Structural identity has been studied in theory and practice over the past decades, but only recently has it been addressed with representational learning techniques. This work presents \textit{struc2vec}, a novel and flexible framework for learning latent representations for the structural identity of nodes. \textit{struc2vec} uses a hierarchy to measure node similarity at different scales, and constructs a multilayer graph to encode structural similarities and generate structural context for nodes. Numerical experiments indicate that state-of-the-art techniques for learning node representations fail in capturing stronger notions of structural identity, while \textit{struc2vec} exhibits much superior performance in this task, as it overcomes limitations of prior approaches. As a consequence, numerical experiments indicate that \textit{struc2vec} improves performance on classification tasks that depend more on structural identity. 
\end{abstract}

%
%




\maketitle

\section{Introduction}
\label{sec:intro}

In almost all networks, nodes tend to have one or more functions that greatly determine their role in the system. For example, individuals in a social network have a social role or social position~\cite{lorrain1971structural,Sailer1978}, while proteins in a protein-protein interaction (PPI) network exert specific functions~\cite{Singh2008,Atias2012}. Intuitively, different nodes in such networks may perform similar functions, such as interns in the social network of a corporation or catalysts in the PPI network of a cell. Thus, nodes can often be partitioned into equivalent classes with respect to their function in the network. 

Although identification of such functions often leverage node and edge attributes, a more challenging and interesting scenario emerges when node function is defined solely by the network structure. In this context, not even the labels of the nodes matter but just their relationship to other nodes (edges). Indeed, mathematical sociologists have worked on this problem since the 1970s, defining and computing {\em structural identity} of individuals in social networks~\cite{lorrain1971structural,Sailer1978,Pizarro2007}. Beyond sociology, the role of webpages in the webgraph is another example of identity (in this case, hubs and authorities) emerging from the network structure, as defined by the celebrated work of Kleinberg~\cite{Kleinberg1999}. 

The most common practical approaches to determine the structural identity of nodes are based on distances or recursions. In the former, a distance function that leverages the neighborhood of the nodes is used to measure the distance between all node pairs, with clustering or matching then performed to place nodes into equivalent classes~\cite{Leicht2006,Fouss2007}. In the later, a recursion with respect to neighboring nodes is constructed and then iteratively unfolded until convergence, with final values used to determine the equivalent classes~\cite{Kleinberg1999,Blondel2004,Zager2008}. While such approaches have advantages and disadvantages, we provide an alternative methodology, one based on unsupervised learning of representations for the structural identity of nodes (to be presented).
\begin{figure}[t]
\includegraphics[width=.5\textwidth]{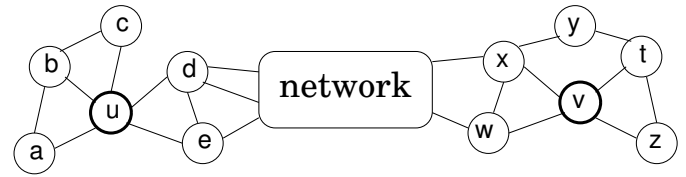}
\caption{An example of two nodes ($u$ and $v$) that are structurally similar (degrees 5 and 4, connected to 3 and 2 triangles, connected to the rest of the network by two nodes), but very far apart in the network.}
\label{fig:example}
\end{figure}
Recent efforts in learning latent representations for nodes in networks have been quite successful in performing classification and prediction tasks~\cite{node2vec-kdd2016,subgraph2vec,Perozzi2014,line}. In particular, these efforts encode nodes using as context a generalized notion of their neighborhood (e.g., $w$ steps of a random walk, or nodes with neighbors in common). In a nutshell, nodes that have neighborhoods with similar sets of nodes should have similar latent representations. But neighborhood is a local concept defined by some notion of proximity in the network. Thus, two nodes with neighborhoods that are structurally similar but that are far apart will not have similar latent representations. Figure \ref{fig:example} illustrates the problem, where nodes $u$ and $v$ play similar roles (i.e., have similar local structures) but are very far apart in the network. Since their neighborhoods have no common nodes, recent approaches cannot capture their structural similarity (as we soon show).

It is worth noting why recent approaches for learning node representations such as \textit{DeepWalk}~\cite{Perozzi2014} and \textit{node2vec}~\cite{node2vec-kdd2016} succeed in classification tasks but tend to fail in structural equivalence tasks. The key point is that many node features in most real networks exhibit a strong homophily (e.g., two blogs with the same political inclination are much more likely to be connected than at random). Neighbors of nodes with a given feature are more likely to have the same feature. Thus, nodes that are ``close'' in the network and in the latent representation will tend to share features. Likewise, two nodes that are ``far'' in the network will tend to be separated in the latent representation, independent of their local structure. Thus, structural equivalence will not properly be captured in the latent representation. However, if classification is performed on features that depend more on structural identity and less on homophily, then such recent approaches are likely to be outperformed by latent representations that better capture structural equivalence (as we soon show). 

Our main contribution is a flexible framework for learning latent representations for the structural identity of nodes, called \textit{struc2vec}. It is an alternative and powerful tool to the study of structural identity through latent representations. The key ideas of \textit{struc2vec} are:
\begin{itemize}
\item
Assess structural similarity between nodes independently of node and edge attributes as well as their position in the network. Thus, two nodes that have a similar local structure will be considered so, independent of network position and node labels in their neighborhoods. Our approach also does not require the network to be connected, and identifies structurally similar nodes in different connected components. 

\item
Establish a hierarchy to measure structural similarity, allowing progressively more stringent notions of what it means to be structurally similar. In particular, at the bottom of the hierarchy, structural similarity between nodes depend only on their degrees, while at the top of the hierarchy similarity depends on the entire network (from the viewpoint of the node). 

\item
Generates random {\em contexts} for nodes, which are sequences of structurally similar nodes as observed by a weighted random walk traversing a multilayer graph (and {\em not} the original network). Thus, two nodes that frequently appear with similar contexts will likely have similar structure. Such context can be leveraged by language models to learn latent representation for the nodes. 
\end{itemize}
We implement an instance of \textit{struc2vec} and show its potential through numerical experiments on toy examples and real networks, comparing its performance with \textit{DeepWalk}~\cite{Perozzi2014} and \textit{node2vec}~\cite{node2vec-kdd2016} -- two state-of-the-art techniques for learning latent representations for nodes, and with \textit{RolX}~\cite{henderson2012rolx} -- a recent approach to identify roles of nodes. Our results indicate that while \textit{DeepWalk} and \textit{node2vec} fail to capture the notion of structural identity,  \textit{struc2vec} excels on this task -- even when the original network is subject to strong random noise (random edge removal). We also show that \textit{struc2vec} is superior in a classification task where node labels depends more on structural identity (i.e., air-traffic networks with labels representing airport activity).

The remainder of this paper is organized as follows. Section~\ref{sec:related} briefly overviews the recent related work on learning latent representations of nodes in networks. Section~\ref{sec:struc2vec} presents the \textit{struc2vec} framework in detail. Experimental evaluation and comparison to other methods are shown in Section~\ref{sec:experiments}. Finally, Section~\ref{sec:conclusion} concludes the paper with a brief discussion.

\section{Related work} \label{sec:related}
Embedding network nodes in (Euclidean) space has received much attention over the past decades from different communities. The technique is instrumental for Machine Learning applications that leverage network data, as node embeddings can be directly used in tasks such as classification and clustering. 

In Natural Language Processing~\cite{nnlm}, generating dense embeddings for sparse data has a long history. Recently, Skip-Gram~\cite{word2vecmiko,skipgram-mikolov} was proposed as an efficient technique to learn embeddings for text data (e.g., sentences). Among other properties,
the learned language model places semantically similar words near each other in space.

Learning a language model from a network was first proposed by \textit{DeepWalk}~\cite{Perozzi2014}. It uses random walks to generate sequences of nodes from the network, which are then treated as sentences by Skip-Gram. Intuitively, nodes close in the network will tend to have similar contexts (sequences) and thus have embeddings that are near one another. This idea was later extended by \textit{node2vec}~\cite{node2vec-kdd2016}. By proposing a biased second order random walk model, \textit{node2vec} provides more flexibility when generating the context of a vertex. In particular, the edge weights driving the biased random walks can be designed in an attempt to capture both vertex homophily and structural equivalence. However, a fundamental limitation is that structurally similar nodes will never share the same context if their distance (hop count) is larger than the Skip-Gram window. 

\textit{subgraph2vec}~\cite{subgraph2vec} is another recent approach for learning embeddings for rooted subgraphs, and unlike the previous techniques it does not use random walks to generate context. Alternatively, the context of a node is simply defined by its neighbors. Additionally, \textit{subgraph2vec} captures structural equivalence by embedding nodes with the same local structure to the same point in space. Nonetheless, the notion of structural equivalence is very rigid since it is defined as a binary property dictated by the Weisfeiler-Lehman isomorphism test~\cite{wlkernel}. Thus, two nodes that are structurally very similar (but fail the test) and have non-overlapping neighbors may not be close in space. 

Similarly to \textit{subgraph2vec}, considerable effort has recently been made on learning richer representations for network nodes~\cite{structural,dnnreps}. However, building representations that explicitly capture structural identity is a relative orthogonal problem that has not received much attention. This is the focus of \textit{struc2vec}.

%

A recent approach to explicitly identify the role of nodes using just the network structure is \textit{RolX}~\cite{henderson2012rolx}. This unsupervised approach is based on enumerating various structural features for nodes, finding the more suited basis vector for this joint feature space, and then assigning for every node a distribution over the identified roles (basis), allowing for mixed membership across the roles. Without explicitly considering node similarity or node context (in terms of structure), \textit{RolX} is likely to miss node pairs that are structurally equivalent (to be shown).


\section{\textit{struc2vec}} \label{sec:struc2vec}
\label{sec:struct2vec}

Consider the problem of learning representations that capture the structural identity of nodes in the network. A successful approach should exhibit two desired properties: 
\begin{itemize}
\item
The distance between the latent representation of nodes should be strongly correlated to their structural similarity. Thus, two nodes that have identical local network structures should have the same latent representation, while nodes with different structural identities should be far apart. 

\item
The latent representation should not depend on any node or edge attribute, including the node labels. Thus, structurally similar nodes should have close latent representation, independent of node and edge attributes in their neighborhood. The structural identity of nodes must be independent of its ``position" in the network. 
\end{itemize}
Given these two properties, we propose \textit{struct2vec}, a general framework for learning latent representations for nodes composed of four main steps, as follows:
\begin{enumerate}
\item
Determine the structural similarity between each vertex pair in the graph for different neighborhood sizes. This induces a hierarchy in the measure for structural similarity between nodes, providing more information to assess structural similarity at each level of the hierarchy.

\item
Construct a weighted multilayer graph where all nodes in the network are present in every layer, and each layer corresponds to a level of the hierarchy in measuring structural similarity. Moreover, edge weights among every node pair within each layer are inversely proportional to their structural similarity.

\item
Use the multilayer graph to generate context for each node. In particular, a biased random walk on the multilayer graph is used to generate node sequences. These sequences are likely to include nodes that are more structurally similar.

\item
Apply a technique to learn latent representation from a context given by the sequence of nodes, for example, Skip-Gram. 
\end{enumerate}
Note that \textit{struct2vec} is quite flexible as it does not mandates any particular structural similarity measure or representational learning framework. In what follows, we explain in detail each step of \textit{struct2vec} and provide a rigorous approach to a hierarchical measure of structural similarity. 

\subsection{Measuring structural similarity}
\label{defining_context}


The first step of \textit{struct2vec} is to determine a structural similarity between two nodes without using any node or edge attributes. Moreover, this similarity metric should be hierarchical and cope with increasing neighborhood sizes, capturing more refined notions of structural similarity. Intuitively, two nodes that have the same degree are structurally similar, but if their neighbors also have the same degree, then they are even more structurally similar. 

Let $G=(V,E)$ denote the undirected, unweighted network under consideration with vertex set $V$ and edge set $E$, where $n=|V|$ denotes the number of nodes in the network and $k^*$ its diameter. Let $R_k(u)$ denote the set of nodes at distance (hop count) exactly $k \geq 0$ from $u$ in $G$. Note that $R_1(u)$ denotes the set of neighbors of $u$ and in general, $R_k(u)$ denotes the ring of nodes at distance $k$. Let $s(S)$ denote the ordered degree sequence of a set $S \subset V$ of nodes.

By comparing the ordered degree sequences of the rings at distance $k$ from both $u$ and $v$ we can impose a hierarchy to measure structural similarity. In particular, let $f_k(u,v)$ denote the {\em structural distance} between $u$ and $v$ when considering their $k$-hop neighborhoods (all nodes at distance less than or equal to $k$ and all edges among them). In particular, we define: 
\begin{equation}
\begin{split}
 f_k(u,v) \, = \, & f_{k-1}(u,v) \, + \, g(s(R_{k}(u)), s(R_{k}(v))), \\
& \,\,\, k\geq 0 \, \mbox{ and } \, |R_{k}(u)|, |R_{k}(v)| > 0
\label{eq:fk}
\end{split}
\end{equation}
where $g(D_1,D_2) \geq 0$ measures the distance between the ordered degree sequences $D_1$ and $D_2$ and $f_{-1} = 0$. Note that by definition $f_k(u,v)$ is non-decreasing in $k$ and is defined only when both $u$ or $v$ have nodes at distance $k$. Moreover, using the ring at distance $k$ in the definition of $f_k(u,v)$ forces the comparison between the degree sequences of nodes that are at the same distance from $u$ and $v$. Finally, note that if the $k$-hop neighborhoods of $u$ and $v$ are isomorphic, and maps $u$ onto $v$, then $f_{k-1}(u,v) = 0$.

A final step is determining the function that compares two degree sequences. Note that $s(R_{k}(u))$ and $s(R_{k}(v))$ can be of different sizes and its elements are arbitrary integers in the range $[0,n-1]$ with possible repetitions. We adopt Dynamic Time Warping (DTW) to measure the distance between two ordered degree sequences, a technique that can cope better with sequences of different sizes and loosely compares sequence patterns~\cite{rakthanmanon2013,salvador2004fastdtw}.

Informally, DTW finds the optimal alignment between two sequences $A$ and $B$. Given a distance function $d(a,b)$ for the elements of the sequence, DTW matches each element $a \in A$ to $b \in B$, such that the sum of the distances between matched elements is minimized. 
Since elements of sequence $A$ and $B$ are degrees of nodes, we adopt the following distance function:
\begin{equation} \label{eq:distdtw}
d(a,b) = \frac{\max(a,b)}{\min(a,b)} - 1
\end{equation}
Note that when $a=b$ then $d(a,b)=0$. Thus, two identical ordered degree sequences will have zero distance. Also note that by taking the ratio between the maximum and the minimum, the degrees 1 and 2 are much more different than degrees 101 and 102, a desired property when measuring the distance between node degrees. Finally, while we use DTW to assess the similarity between two ordered degree sequences, any other cost function could be adopted by our framework. 




\subsection{Constructing the context graph}
\label{sec:distancegraph}

We construct a multilayer weighted graph that encodes the structural similarity between nodes. Recall that $G=(V,E)$ denotes the original network (possibly not connected) and $k^*$ its diameter. Let $M$ denote the multilayer graph where layer $k$ is defined using the $k$-hop neighborhoods of the nodes.

Each layer $k = 0, \ldots, k^*$ is formed by a weighted undirected complete graph with node set $V$, and thus, ${n \choose 2}$ edges. The edge weight between two nodes in a layer is given by: 
\begin{equation} \label{eq:distweight}
w_k(u,v) = e^{-f_k(u,v)}  , \;\;  k = 0, \ldots, k^*
\end{equation}
Note that edges are defined only if $f_k(u,v)$ is defined and that weights are inversely proportional to structural distance, and assume values smaller than or equal to 1, being equal to 1 only if $f_k(u,v) = 0$. Note that nodes that are structurally similar to $u$ will have larger weights across various layers of $M$. 

We connect the layers using directed edges as follows. Each vertex is connected to its corresponding vertex in the layer above and below (layer permitting). Thus, every vertex $u \in V$ in layer $k$ is connected to the corresponding vertex $u$ in layer $k+1$ and $k-1$. The edge weight between layers are as follows:
\begin{equation} \label{eq:distweight}
\begin{split}
&w(u_k,u_{k+1}) = \log(\Gamma_k(u) + e) , \;\;  k=0, \ldots, k^*-1 \\
&w(u_k,u_{k-1}) = 1 , \;\;  k=1, \ldots, k^*
\end{split}
\end{equation}
where $\Gamma_k(u)$ is number of edges incident to $u$ that have weight larger than the average edge weight of the complete graph in layer $k$. In particular:
\begin{equation} \label{eq:gamma}
\Gamma_k(u) = \sum_{v\in V} \mathds{1}(w_k(u,v) > \overline{w_k})
\end{equation}
where $\overline{w_k} = \sum_{(u,v) \in {V \choose 2}} w_k(u,v) / {n \choose 2}$. Thus, $\Gamma_k(u)$ measures the similarity of node $u$ to other nodes in layer $k$. Note that if $u$ has many similar nodes in the current layer, then it should change layers to obtain a more refined context. Note that by moving up one layer the number of similar nodes can only decrease. Last, the $\log$ function simply reduces the magnitude of the potentially large number of nodes that are similar to $u$ in a given layer. 

Finally, note that $M$ has $n k^*$ vertices and at most $k^*{n \choose 2}+2n(k^*-1)$ weighted edges. In Section~\ref{sec:complexity} we discuss how to reduce the complexity of generating and storing $M$. 

\subsection{Generating context for nodes} \label{seqs}

The multilayer graph $M$ is used to generate structural context for each node $u \in V$. Note that $M$ captures the structure of structural similarities between nodes in $G$ using absolutely no label information. As in previous works, \textit{struct2vec} uses random walks to generate sequence of nodes to determine the context of a given node. In particular, we consider a biased random walk that moves around $M$ making random choices according to the weights of $M$. Before each step, the random walk first decides if it will change layers or walk on the current layer (with probability $q>0$ the random walk stays in the current layer). 

Given that it will stay in the current layer, the probability of stepping from node $u$ to node $v$ in layer $k$ is given by: 
\begin{equation} \label{eq:edgeprob}
p_k(u,v) = \frac{e^{-f_k(u,v)}}{Z_k(u)} 
\end{equation}
where $Z_k(u)$ is the normalization factor for vertex $u$ in layer $k$, simply given by:
\begin{equation} \label{eq:partition}
Z_k(u) = \sum_{\substack{{v \in V} \\ {v \neq u}}} e^{-f_k(u,v)} 
\end{equation}
Note that the random walk will prefer to step onto nodes that are structurally more similar to the current vertex, avoiding nodes that have very little structural similarity with it. Thus, the context of a node $u \in V$ is likely to have structurally similar nodes, independent of their labels and position in the original network $G$. 

With probability $1-q$, the random walk decides to change layers, and moves to the corresponding node either in layer $k+1$ or layer $k-1$ (layer permitting) with probability proportional to the edge weights. In particular:
\begin{equation} \label{eq:layerprob}
\begin{split}
&p_k(u_k,u_{k+1}) = \frac{w(u_k,u_{k+1})}{w(u_k,u_{k+1})+w(u_k,u_{k-1})} \\
&p_k(u_k,u_{k-1}) = 1 - p_k(u_k,u_{k+1})
\end{split}
\end{equation}
Note that every time the walker steps within a layer it includes the current vertex as part of its context, independent of the layer. Thus, a vertex $u$ may have a given context in layer $k$ (determined by the structural similarity of this layer), but have a subset of this context at layer $k+1$, as the structural similarity cannot increase as we move to higher layers. This notion of a hierarchical context across the layers is a fundamental aspect of the proposed methodology. 

Finally, for each node $u \in V$, we start a random walk in its corresponding vertex in layer 0. Random walks have a fixed and relatively short length (number of steps), and the process is repeated a certain number of times, giving rise to multiple independent walks (i.e., the multiple contexts of node $u$). 

\subsection{Learning a language model} \label{skip-gram}

Recent language modeling techniques have been extensively used to learn word embeddings, and only require sets of sentences in order to generate meaningful representations. Informally, the task can be defined as learning word probabilities given a context.
In particular, Skip-Gram~\cite{skipgram-mikolov} has proven to be effective at learning meaningful representations for a variety of data. In order to apply it to networks, it suffices to use artificially generated node sequences instead of word sentences. In our framework, these sequences are generated by biased random walks on the multilayer graph $M$. 
Given a node, Skip-Gram aims to maximize the likelihood of its context in a sequence, where a node's context is given by a window of size $w$ centered on it. 


For this work we use Hierarchical Softmax, where 
conditional symbol probabilities are calculated using a tree of binary classifiers. For each node $v_j \in V$, Hierarchical Softmax assigns a specific path in the classification tree, defined by a set of tree nodes $n(v_j,1), n(v_j,2), \dots, n(v_j,h)$, where $n(v_j,h)=v_j$. In this setting, we have:
\begin{equation} \label{eq:hs}
P(v_j| v_i) = \prod_{k=1}^h C(n(v_j,k), v_i)
\end{equation}
where $C$ is a binary classifier present in every node in the tree. Note that since Hierarchical Softmax operates on a binary tree, we have that $h = O(\log |V|)$.

We train Skip-Gram according to its optimization problem given by equation (\ref{eq:hs}). Note that while we use Skip-Gram to learn node embeddings, any other technique to learn latent representations for text data could be used in our framework. 

\subsection{Complexity and optimizations}
\label{sec:complexity}

In order to construct $M$, the structural distance between every node pair for every layer must be computed, namely, $f_k(u,v)$ for $u,v \in V$, and $0 \leq k \leq k^*$. However, $f_k(u,v)$ uses the result of a DTW calculation between two degree sequences. While classic implementation of DTW has complexity $O(\ell^2)$, fast techniques have complexity $O(\ell)$, where $\ell$ is the size of the largest sequence~\cite{salvador2004fastdtw}. Let $d_{\max}$ denote the largest degree in the network. Then, the size of the degree sequence $|s(R_{k}(u))| \leq \min(d_{\max}^k, n)$, for any node $u$ and layer $k$. Since in each layer there are ${n \choose 2}$ pairs, the complexity of computing all distances for layer $k$ is $O(n^2 \min(d_{\max}^k, n))$. The final complexity is then $O(k^* n^3)$. In what follows we describe a series of optimizations that will significantly reduce the computation and memory requirements of the framework.

\paragraph{Reducing the length of degree sequences (OPT1).} Although degree sequences at layer $k$ have lengths bounded by $\min(d_{\max}^k, n)$, for some networks this can be quite large even for small $k$ (e.g., for $k=3$ the sequences are already $O(n)$). To reduce the cost of comparing large sequences, we propose compressing the ordered degree sequence as follows. For each degree in the sequence, we count the number of occurrences of that degree. The compressed ordered degree sequence is a tuple with the degree and the number of occurrences. Since many nodes in a network tend to have the same degree, in practice the compressed ordered degree sequence can be an order of magnitude smaller than the original. 
%

Let $A'$ and $B'$ denote the compressed degree sequences of $A$ and $B$, respectively. Since the elements of $A'$ and $B'$ are tuples, we adapt the DTW pairwise distance function as follows:
\begin{equation} \label{eq:distdtwvector}
\mathrm{dist}(\bm{a},\bm{b}) = \bigg( \frac{\max(a_0,b_0)}{\min(a_0,b_0)} - 1 \bigg) \max(a_1,b_1)
\end{equation}
where $\bm{a} = (a_0,a_1)$ and $\bm{b} = (b_0,b_1)$ are tuples in $A'$ and $B'$, respectively; $a_0$ and $b_0$ are the degrees; $a_1$ and $b_1$ are the number of occurrences. 
Note that using the compressed degree sequence leads to comparisons between pieces of the original sequences that have the same degree (as opposed to comparing every degree). Thus, equation~(\ref{eq:distdtwvector}) leads to an approximation of the DTW on the original degree sequences, as given by equation~(\ref{eq:distdtw}). However, DTW now operates on $A'$ and $B'$, which are much shorter than $A$ and $B$, respectively. 


\paragraph{Reducing the number of pairwise similarity calculations (OPT2).}
While the original framework assesses the similarity between every node pair at every layer $k$, clearly this seems unnecessary. Consider two nodes with very different degrees (eg., 2 and 20). Their structural distance even for $k=0$ will be large, and consequently the edge between them in $M$ will have a very small weight. Thus, when generating context for these nodes, the random walk is unlikely to traverse this edge. Consequently, not having this edge in $M$ will not significantly change the model. 

We limit the number of pairwise similarity calculations to $\Theta(\log n)$ per node, for every level $k$. Let $J_u$ denote the set of nodes that will be neighbors of $u$ in $M$, which will be the same for every level. $J_u$ should have the nodes most structurally similar to $u$. In order to determine $J_u$, we take the nodes that have degrees most similar to $u$. This can be computed efficiently by performing a binary search on the ordered degree sequence of all nodes in the network (for the degree of node $u$), and taking $\log n$ consecutive nodes on each direction after the search completes. Thus, computing $J_u$ has complexity $\Theta(\log n)$. Computing $J_u$ for all nodes has complexity $\Theta(n \log n)$ which is also needed for sorting the degrees of the network. As for memory requirements, each layer of $M$ will now have $\Theta(n \log n)$ edges as opposed to $\Theta(n^2)$.

\paragraph{Reducing the number of layers (OPT3).}
The number of layers in $M$ is given by the diameter of the network, $k^*$. However, for many networks the diameter can be much larger than the average distance. Moreover, the importance of assessing the structural similarity between two nodes diminishes with arbitrarily large values for $k$. In particular, when $k$ is near $k^*$ the length of the degree sequences of the rings become relatively short, and thus $f_k(u,v)$ is not much different from $f_{k-1}(u,v)$. Therefore, we cap the number the layers in $M$ to a fixed constant $k' < k^*$, capturing the most important layers for assessing structural similarity. This significantly reduces the computational and memory requirements for constructing $M$. 


Although the combination of the above optimizations affects the capacity of the framework in generating good representations for nodes that are structurally similar, we will show that their impact is marginal and sometimes even beneficial. Thus, the benefits in reducing computational and memory requirements of the framework greatly outweighs its drawbacks. Last, we make \textit{struc2vec} available at: \url{https://github.com/leoribeiro/struc2vec}

\section{Experimental Evaluation} \label{sec:experiments}

In what follows we evaluate \textit{struct2vec} in different scenarios in order to illustrate its potential in capturing the structural identity of nodes, also in light of state-of-the-art techniques for learning node representations.  

\subsection{Barbell graph}


We denote $B(h,k)$ as the $(h,k)$-barbell graph which consists of two complete graphs $K_1$ and $K_2$ (each having $h$ nodes) connected by a path graph $P$ of length $k$. Two nodes $b_1 \in V(K_1)$ and $b_2 \in V(K_2)$ act as the bridges. Using $\{ p_1, \dots, p_k \}$ to denote $V(P)$, we connect $b_1$ to $p_1$ and $b_2$ to $p_k$, thus connecting the three graphs.

\begin{figure*}[t]
\includegraphics[width=1\textwidth]{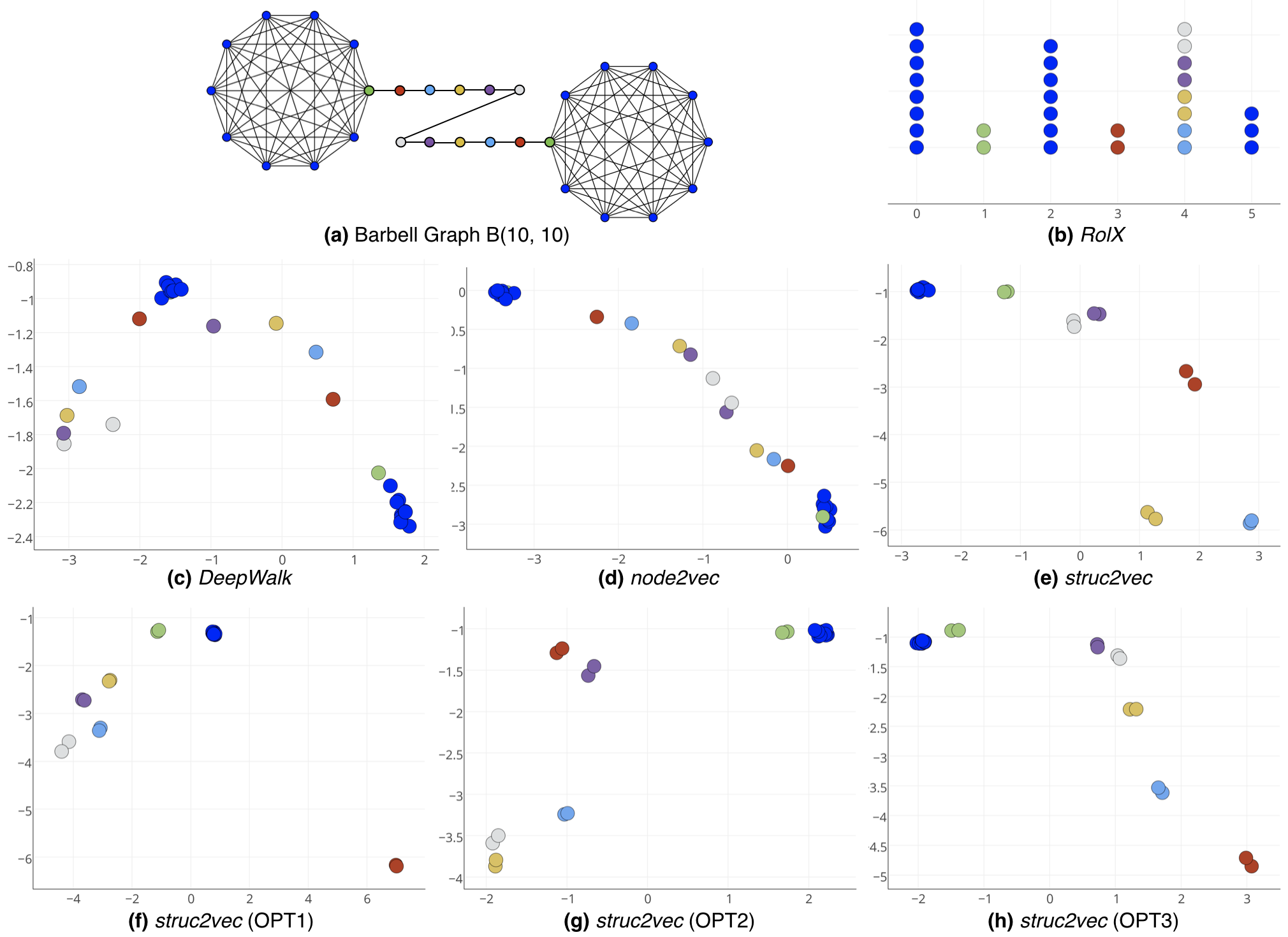}
\caption{(a) Barbell graph $B(10,10)$. (b) Roles identified by~\textit{RolX}. Latent representations in $\mathbb{R}^2$ learned by (c)~DeepWalk, (d)~\textit{node2vec} and (e,f,g,h)~\textit{struc2vec}. Parameters used for all methods: number of walks per node: 20, walk length: 80, skip-gram window size: 5. For \textit{node2vec}: $p=1$ and $q=2$.}
\label{barbell_comparison}
\end{figure*}

The barbell graph has a significant number of nodes with the same structural identity. Let $C_1 = V(K_1) \setminus \{b_1\}$ and $C_2 = V(K_2) \setminus \{b_2\}$. Note that all nodes $v \in \{C_1 \cup C_2\}$ are structurally equivalent, in the strong sense that there exists an automorphism between any pair of such nodes. Additionally, we also have that all node pairs $\{p_i, p_{k-i}\}$, for $1 \leq i \leq k-1$, along with the pair $\{b_1,b_2\}$, are structurally equivalent in the same strong sense. Figure~\ref{barbell_comparison}a illustrates a $B(10,10)$ graph, where structurally equivalent nodes have the same color.

Thus, we expect \textit{struct2vec} to learn vertex representations that capture the structural equivalence mentioned above. Every node pair that is structurally equivalent should have similar latent representation. Moreover, the learned representations should also capture structural hierarchies: while the node $p_1$ is not equivalent to neither nodes $p_2$ or $b_1$, we can clearly see that from a structural point of view it is more similar to $p_2$ (it suffices to compare their degrees).

Figure~\ref{barbell_comparison} shows the latent representations learned by \textit{DeepWalk}, \textit{node2vec} and \textit{struct2vec} for $B(10,10)$. \textit{DeepWalk} fails to capture structural equivalences, which is expected since it was not designed to consider structural identities.
As illustrated, \textit{node2vec} does not capture structural identities even with different variations of its parameters $p$ and $q$. In fact, it learns mostly graph distances, placing closer in the latent space nodes that are closer (in hops) in the graph. Another limitation of \textit{node2vec} is that Skip-Gram's window size makes it impossible for nodes in $K_1$ and $K_2$ to appear in the same context.

\textit{struct2vec}, on the other hand, learns representations that properly separate the equivalent classes, placing structurally equivalent nodes near one another in the latent space. Note that nodes of the same color are tightly grouped together. Moreover, $p_1$ and $p_{10}$ are placed close to representations for nodes in $K_1$ and $K_2$, as they are the bridges. Finally, note that none of the three optimizations have any significant effect on the quality of the representations. In fact, structurally equivalent nodes are even closer to one another in the latent representations under OPT1. 

Last, we apply \textit{RolX} to the barbell graph (results in Figure~\ref{barbell_comparison}(b)). A total of six roles were identified and some roles indeed precisely captured structural equivalence (roles 1 and 3). However, structurally equivalent nodes (in $K_1$ and $K_2$) were placed in three different roles (role 0, 2, and 5) while role 4 contains all remaining nodes in the path. Thus, although \textit{RolX} does capture some notion of structural equivalence when assigning roles to nodes, \textit{struct2vec} better identifies and separates structural equivalence. 

\subsection{Karate network}

\begin{figure}
\includegraphics[width=.50\textwidth]{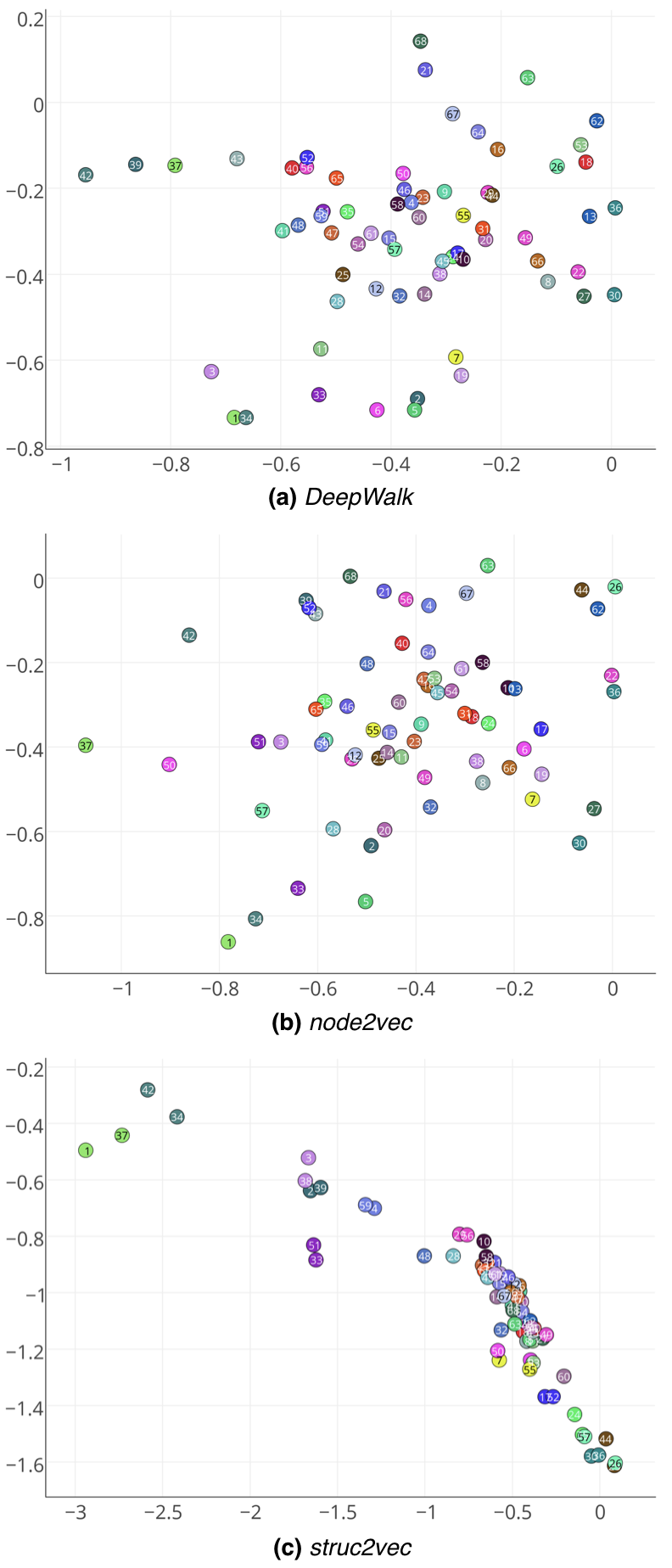}
\caption{Node representations for the mirrored Karate network created by (a)~\textit{DeepWalk}, (b)~\textit{node2vec} and (c)~\textit{struc2vec}. Parameters used for all methods: number of walks per node: 5, walk length: 15, Skip-Gram window size: 3. For \textit{node2vec}: $p=1$ and $q=2$.}
\label{karate-embeddings}
\end{figure}

\begin{figure}
\includegraphics[width=.55\textwidth]{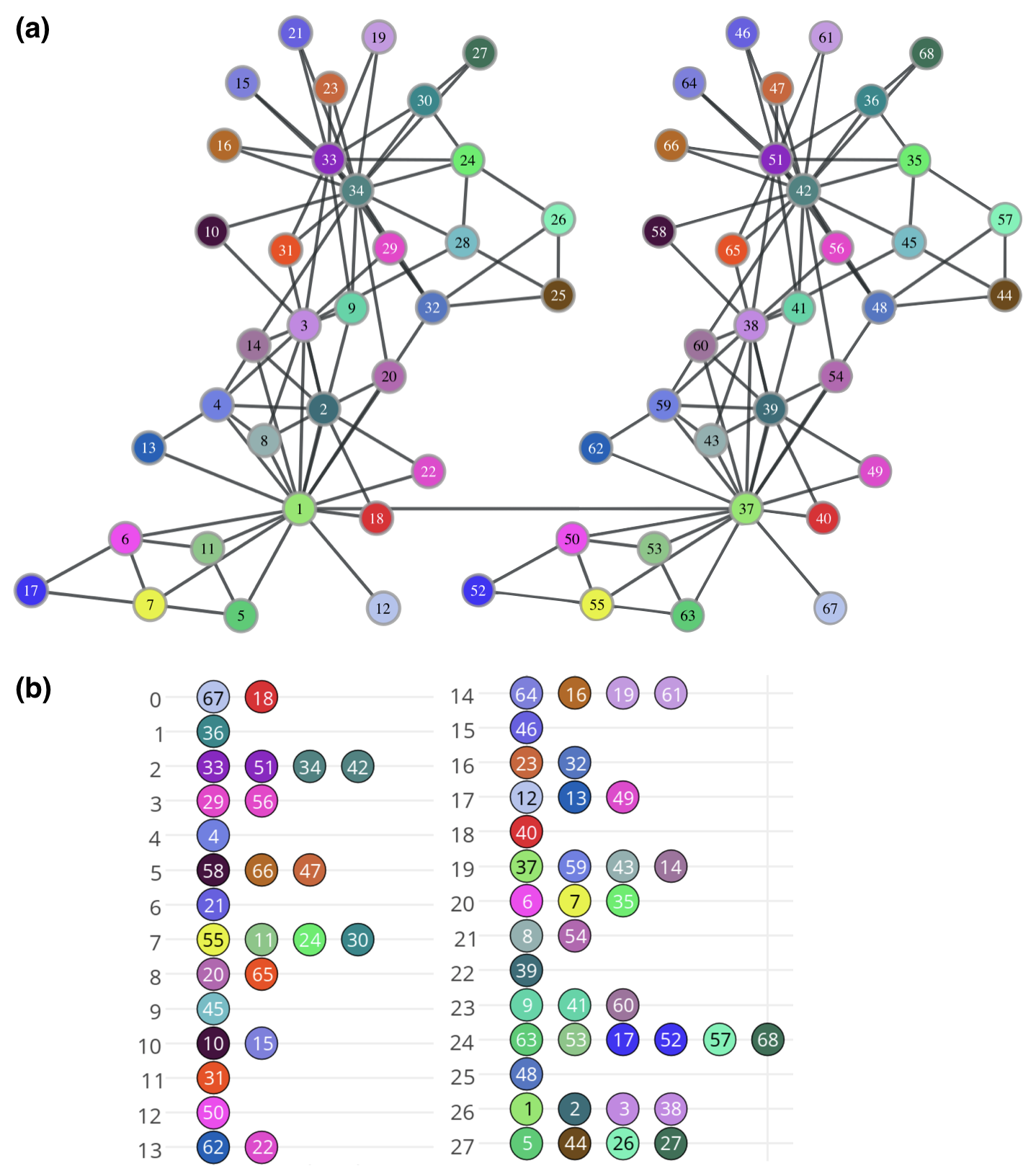} 
\caption{(a) Mirrored Karate network. Identical colors correspond to mirrored nodes. (b) Roles identified by~\textit{RolX}}
\label{karate-graph}
\end{figure}

The Zachary's Karate Club~\cite{zachary1977information} is a network composed of 34 nodes and 78 edges, where each node represents a club member and edges denote if two members have interacted outside the club. In this network, edges are commonly interpreted as indications of friendship between members.

We construct a network composed of two copies $G_1$ and $G_2$ of the Karate Club network, where each node $v \in V(G_1)$ has a mirror node $u \in V(G_2)$. We also connect the two networks by adding an edge between mirrored node pairs $1$ and $37$. Although this is not necessary for our framework, \textit{DeepWalk} and \textit{node2vec} cannot place in the same context nodes in different connected components of the graph. Thus, we add the edge for a more fair comparison with the two baselines. Figure~\ref{karate-graph}a shows mirrored network with corresponding pairs having the same color.

Figure~\ref{karate-embeddings} shows the representations learned by \textit{DeepWalk}, \textit{node2vec} and \textit{struct2vec}. Clearly, \textit{Deepwalk} and \textit{node2vec} fail to group in the latent space structurally equivalent nodes, including mirrored nodes. 

Once again, \textit{struct2vec} manages to learn features that properly capture the structural identity of nodes. Mirrored pairs -- that is, nodes with the same color -- stay close together in the latent space, and there is a complex structural hierarchy in the way the representations are grouped together.

As an example, note that nodes $1$, $34$ and their correspondent mirrors ($37$ and $42$) are in a separate cluster in the latent space. Interestingly, these are exactly the nodes that represent the club instructor Mr. Hi and his administrator John A. The network was gathered after a conflict between the two split the members of the club into two groups -- centered on either Mr. Hi or John A. Therefore, nodes $1$ and $34$ have the specific and similar role of leaders in the network. Note that \textit{struct2vec} captures their function even though there is no edge between them.

Another visible cluster in the latent space is composed of nodes $2, 3, 4$ and $33$, also along with their mirrors. These nodes also have a specific structural identity in the network: all of them have high degrees and are also connected to at least one of the leaders. Lastly, nodes $26$ and $25$ (far right in the latent space) have extremely close representations, which agrees with their structural role: both have low degree and are 2 hops away from leader $34$.

\textit{struct2vec} also captures non-trivial structural equivalences. Note that nodes $7$ and $50$ (pink and yellow) are mapped to close points in the latent space. Surprisingly, these two nodes are structurally equivalent -- there exists an automorphism in the graph that maps one into the other. This can be more easily seen once we note that nodes $6$ and $7$ are also structurally equivalent, and $50$ is the mirrored version of node $6$ (therefore also structurally equivalent). 


Last, Figure~\ref{karate-graph}b shows the roles identified by \textit{RolX} in the mirrored Karate network (28 roles were identified). Note that leaders $1$ and $34$ were placed in different roles. The mirror for $1$ (node $37$) was also placed in a different role, while the mirror for $34$ (node $42$) was placed in the same role as $34$. A total of 7 corresponding pairs (out of 34) were placed in the same role. However, some other structural similarities were also identified -- e.g., nodes $6$ and $7$ are structurally equivalent and were assigned the same role. Again, \textit{RolX} seems to capture some notion of structural similarities among network nodes but \textit{struct2vec} can better identify and separate structural equivalences using latent representations. 



\begin{figure}
\includegraphics[width=.48\textwidth]{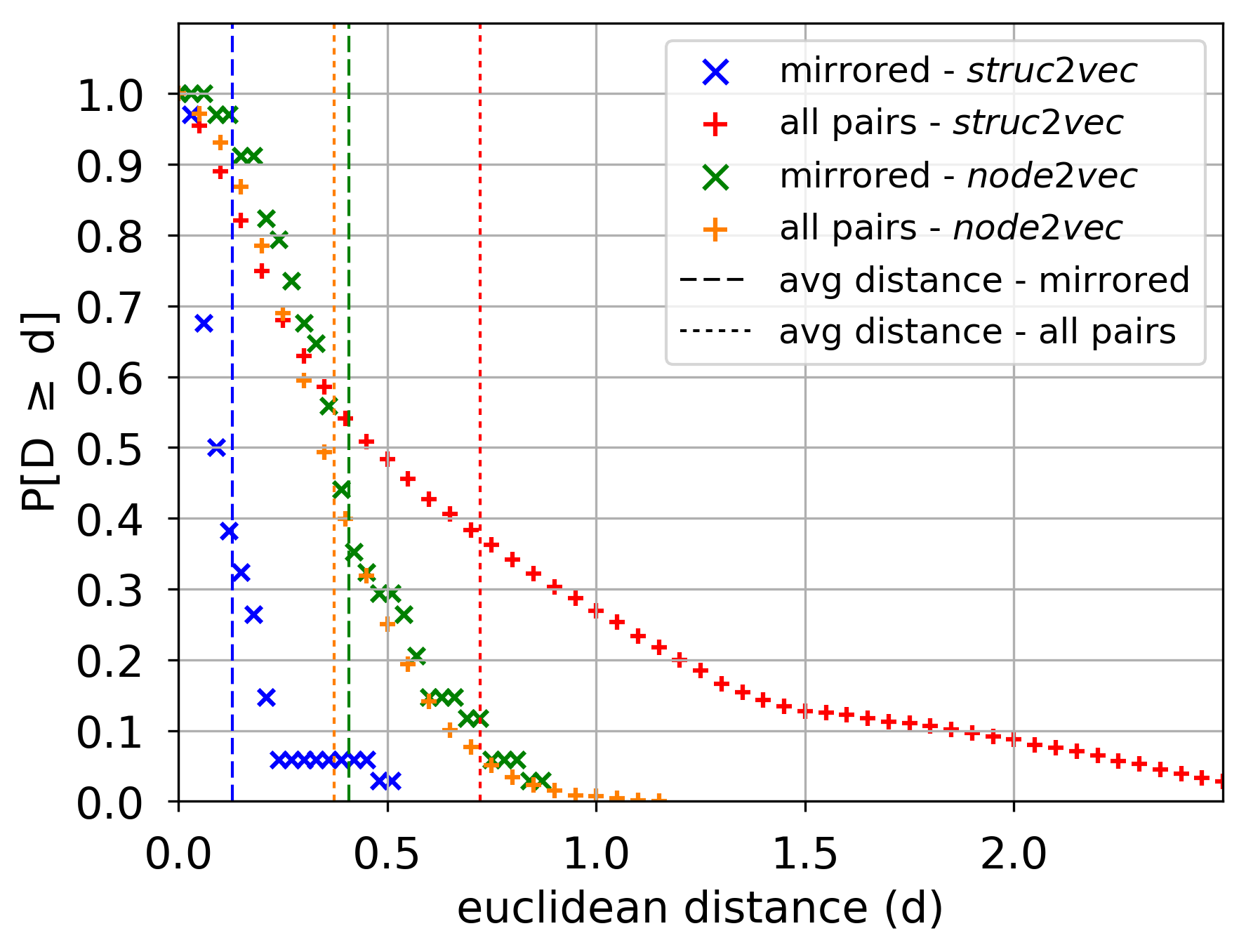} 
\caption{Distance distributions between node pairs (mirrored pairs and all pairs) in the latent space, for the mirrored Karate network learned by \textit{node2vec} and \textit{struc2vec} (as shown in Figure~\ref{karate-embeddings}). Curves marked with $\times$ correspond to distances between mirrored pairs while $+$ corresponds to all pairs; corresponding averages indicated by vertical lines.}
\label{karate-ccdf}
\end{figure}

Consider the distance between the latent representation for nodes. We measure the distance distribution between pairs corresponding to mirrored nodes and among all node pairs (using the representation shown in Figure~\ref{karate-embeddings}). Figure~\ref{karate-ccdf} shows the two distance distributions for the representations learned by \textit{node2vec} and \textit{struc2vec}. For \textit{node2vec} the two distributions are practically identical, indicating that distances between mirrored pairs blend well with all pairs. In contrast, \textit{struc2vec} exhibits two very different distributions: 94\% of mirrored node pairs have distance smaller than 0.25 while 68\% of all node pairs have distance larger than 0.25. Moreover, the average distance between all node pairs is 5.6 times larger than that of mirrored pairs, while this ratio is about slightly {\em smaller} than 1 for \textit{node2vec}. 

To better characterize the relationship between structural distance and distances in the latent representation learned by \textit{struc2vec}, we compute the correlation between the two distances for all node pairs. In particular, for each layer $k$ we compute the Spearman and Pearson correlation coefficients between $f_k(u,v)$, as given by equation~(\ref{eq:fk}), and the euclidean distance between $u$ and $v$ in the learned representation. Results shown in Table~\ref{tab:distcorr} for the mirrored Karate network indeed corroborate that there is a very strong correlation between the two distances, for every layer, captured by both coefficients. This suggests that \textit{struc2vec} indeed captures in the latent space the measure for structural similarity adopted by the methodology. 

\begin{table}[hbtp]
\caption{Pearson and Spearman correlation coefficients between structural distance and euclidean distance in latent space for all node pairs in the mirrored Karate network. }
\begin{center}
\begin{adjustbox}{max width=\textwidth}

\setlength\extrarowheight{1.5pt}
\begin{tabular}{|C{1.3cm}|C{2.5cm}|C{2.8cm}|}
  \hline 
 \myrowcolour%
  Layer & Pearson (p-value) & Spearman (p-value)   \\
    \hline 
      0  & 0.83 (0.0) & 0.74 (0.0)\\
      \myrowcolour%
  \hline 
%
     2  & 0.71 (0.0) & 0.65 (0.0)\\
  \hline
%
       4  & 0.70 (0.0)  & 0.57 (0.0)\\
       \myrowcolour%
  \hline
%
         6  &  0.74 (0.0) & 0.57 (2.37) \\
         \myrowcolour%
  \hline
%
\end{tabular}
\end{adjustbox}
\end{center}
\label{tab:distcorr}
\end{table}

\subsection{Robustness to edge removal}
\label{edgeremoval}

We illustrate the potential of the framework in effectively representing structural identity in the presence of noise. In particular, we randomly remove edges from the network, directly changing its structure. We adopt the parsimonious {\em edge sampling model} to instantiate two structurally correlated 
networks~\cite{pedarsani2011}. 

\begin{figure}[t]
\includegraphics[width=.48\textwidth]{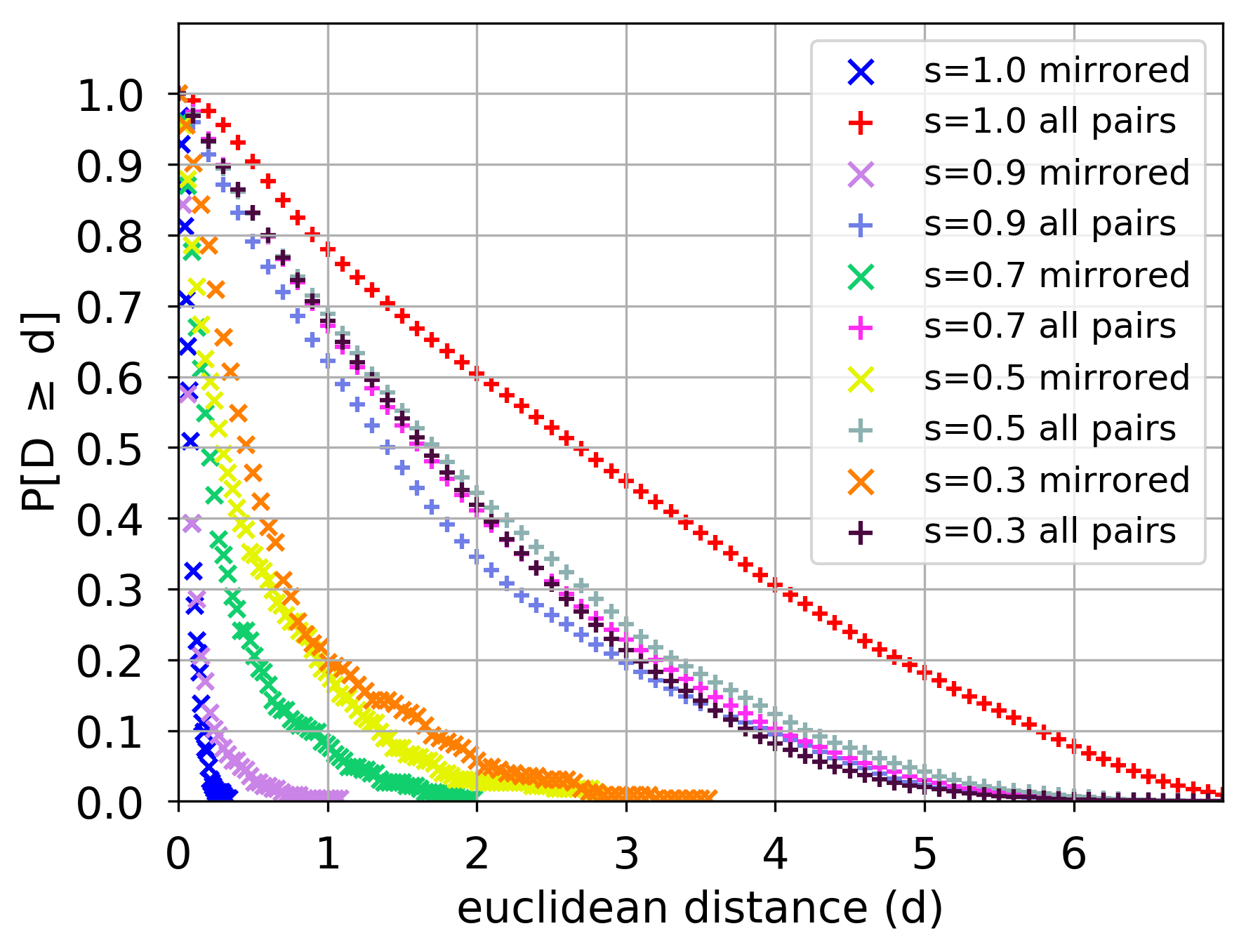} 
\caption{Distance distribution between node pairs in latent space representation (2 dimensions) under the edge sampling model (different values for $s$). Bottom curves (marked with $\times$) are for corresponding node pairs; top curves (marked with $+$) are for all node pairs.}
\label{facebook-ccdf-d2}
\end{figure}

The model works by taking a fixed graph $G=(V,E)$ and generating a graph $G_1$ by sampling each edge $e \in E$ with probability $s$, independently. Thus, each edge of $G$ is present in $G_1$ with probability $s$. Repeat the process again using $G$ to generate another graph $G_2$. Thus, $G_1$ and $G_2$ are structurally correlated through $G$, and $s$ controls the amount of structural correlation. Note that when $s=1$, $G_1$ and $G_2$ are isomorphic, while when $s=0$ all structural identity is lost. 

We apply the edge sampling model to an {\em egonet} extracted from Facebook (224 nodes, 3192 edges, max degree 99, min degree 1)~\cite{leskovec2012learning} to generate $G_1$ and $G_2$ with different values for $s$. We relabel the nodes in $G_2$ (to avoid identical labels) and consider the union of the two graphs as the input network to our framework. Note that this graph has at least two connected components (corresponding to $G_1$ and $G_2$) and every node (in $G_1$) has a corresponding pair (in $G_2$).

Figure~\ref{facebook-ccdf-d2} shows the distance (latent space with 2 dimensions) distribution between corresponding node pairs and all node pairs for various values for $s$ (corresponding averages are shown in Table~\ref{tab:egodistances}). 
For $s=1$, the two distance distributions are strikingly different, with the average distance for all pairs being 21 times larger than that for corresponding pairs. More interestingly, when $s=0.9$ the two distributions are still very different. Note that while further decreasing $s$ does not significantly affect the distance distribution of all pairs, it slowly increases the distribution of corresponding pairs. However, even when $s=0.3$ (which means that the probability that an original edge appears both in $G_1$ and $G_2$ is 0.09, $s^2$), the framework still places corresponding nodes closer in the latent space.

This experiment indicates the robustness of the framework in uncovering the structural identity of nodes even in the presence of structural noise, modeled here through edge removals. 

\begin{table}[hbtp]
\caption{Average and standard deviation for distances between node pairs in the latent space representation (see corresponding distributions in Figure~\ref{facebook-ccdf-d2}).}
\begin{center}
\begin{adjustbox}{max width=\textwidth}
\setlength\tabcolsep{1.5pt}
\setlength\extrarowheight{1pt}
\begin{tabular}{|c|C{3.8cm}|C{3cm}|}
  \hline 
 \myrowcolour%
 s& Corresponding - avg~(std)& All nodes - avg~(std)\\
  \hline
 1.0 & 0.083 (0.05) & 1.780 (1.354) \\
  \hline
  \myrowcolour%
 0.9 & 0.117 (0.142) & 1.769 (1.395)   \\
  \hline
%
 0.3 & 0.674 (0.662)  & 1.962 (1.445)  \\
    \hline
\end{tabular}
\end{adjustbox}
\end{center}
\label{tab:egodistances}
\end{table}

\subsection{Classification}
\label{multiclass}

\begin{figure}
\centering
\includegraphics[width=.48\textwidth]{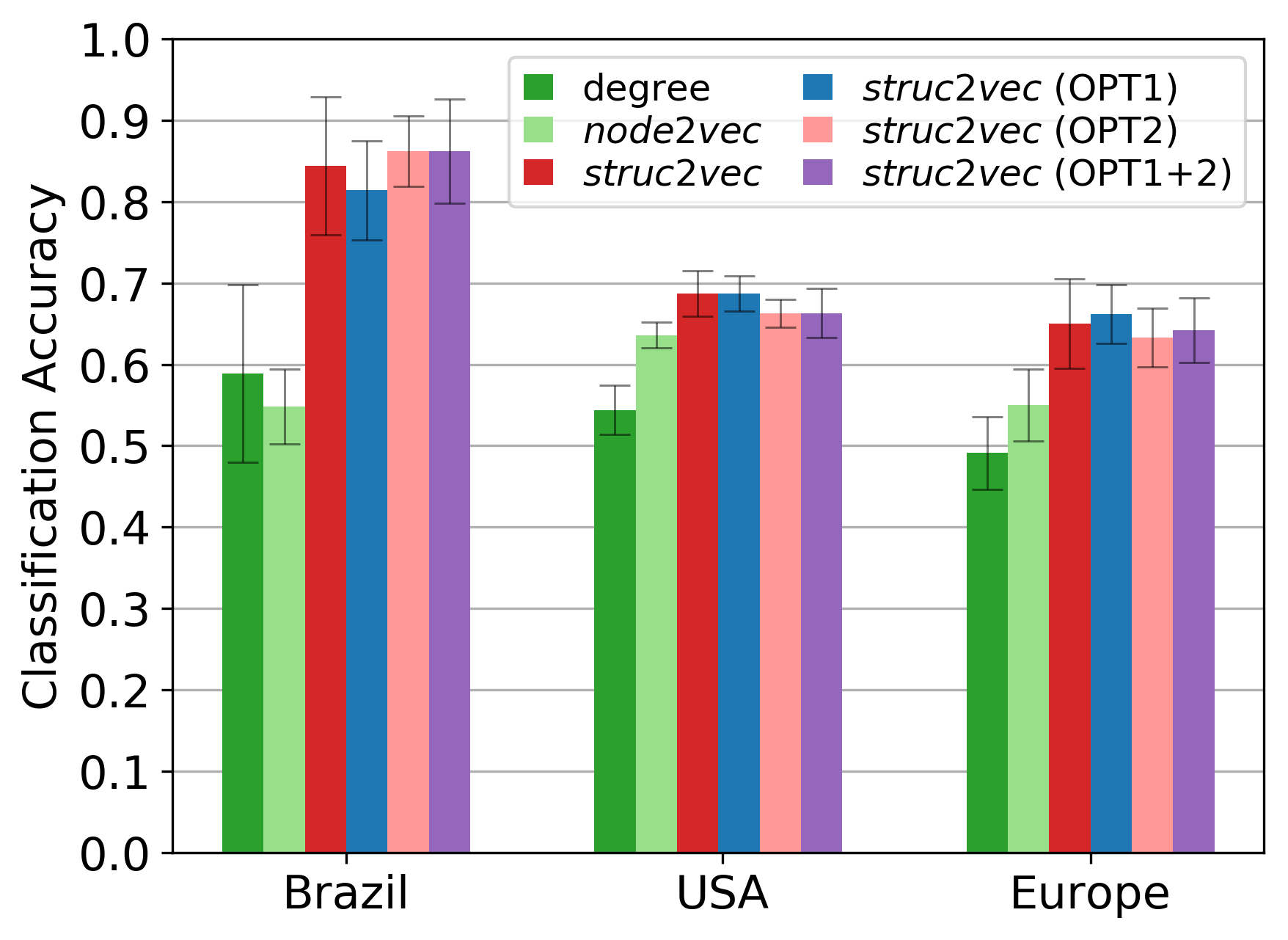} 
\caption[]{Average accuracy for multi-class node classification in air-traffic networks of Brazil, USA and Europe for different node features used in supervised learning.}
\label{classification_accuracy}
\end{figure}

A common application of latent representations for network nodes is classification. \textit{struc2vec} can be leveraged for this task when labels for nodes are more related to their structural identity than to the labels of their neighbors. To illustrate this potential, we consider air-traffic networks: unweighted, undirected networks where nodes correspond to airports and edges indicate the existence of commercial flights. Airports will be assigned a label corresponding to their level of activity, measured in flights or people (discussed below). We consider the following datasets (collected for this study):

\begin{itemize}
    \item \emph{Brazilian air-traffic network:} Data collected from the National Civil Aviation Agency (ANAC)\footnote{http://www.anac.gov.br/} from January to December 2016. The network has 131 nodes, 1,038 edges (diameter is 5). Airport activity is measured by the total number of landings plus takeoffs in the corresponding year. 
    \item \emph{American air-traffic network:} Data collected from the Bureau of Transportation Statistics\footnote{https://transtats.bts.gov/} from January to October, 2016. The network has 1,190 nodes, 13,599 edges (diameter is 8). Airport activity is measured by the total number of people that passed (arrived plus departed) the airport in the corresponding period. 
    \item \emph{European air-traffic network:} Data collected from the Statistical Office of the European Union (Eurostat)\footnote{http://ec.europa.eu/} from January to November 2016. The network has 399 nodes, 5,995 edges (diameter is 5). Airport activity is measured by the total number of landings plus takeoffs in the corresponding period. 
\end{itemize}
For each airport, we assign one of four possible labels corresponding to their activity. In particular, for each dataset, we use the quartiles obtained from the empirical activity distribution to split the dataset in four groups, assigning a different label for each group. Thus, label 1 is given to the 25\% less active airports, and so on. Note that all classes (labels) have the same size (number of airports). Moreover, classes are related more to the role played by the airport. 

We learn latent representations for nodes of each air-traffic network using \textit{struc2vec} and \textit{node2vec} using a grid search to select the best hyperparameters for each case. Note that this step does not use any node label information. The latent representation for each node becomes the feature that is then used to train a supervised classifier (one-vs-rest logistic regression with L2 regularization). We also consider just the node degree as a feature since it captures a very basic notion of structural identity. Last, since classes have identical sizes, we use just the accuracy to assess performance. Experiments are repeated 10 times using random samples to train the classifier (80\% of the nodes used for training) and we report on the average performance.

Figure~\ref{classification_accuracy} shows the classification performance of the different features for all air-traffic networks. Clearly, \textit{struc2vec} outperforms the other approaches, and its optimizations have little influence. For the Brazilian network, \textit{struc2vec} improves classification accuracy by 50\% in comparison to \textit{node2vec}. Interestingly, for this network \textit{node2vec} has average performance (slightly) inferior to node degree, indicating the importance played by the structural identity of the nodes in classification.

\subsection{Scalability}
\label{sec:scalability}

\begin{figure}[h]
\centering
\includegraphics[width=.46\textwidth]{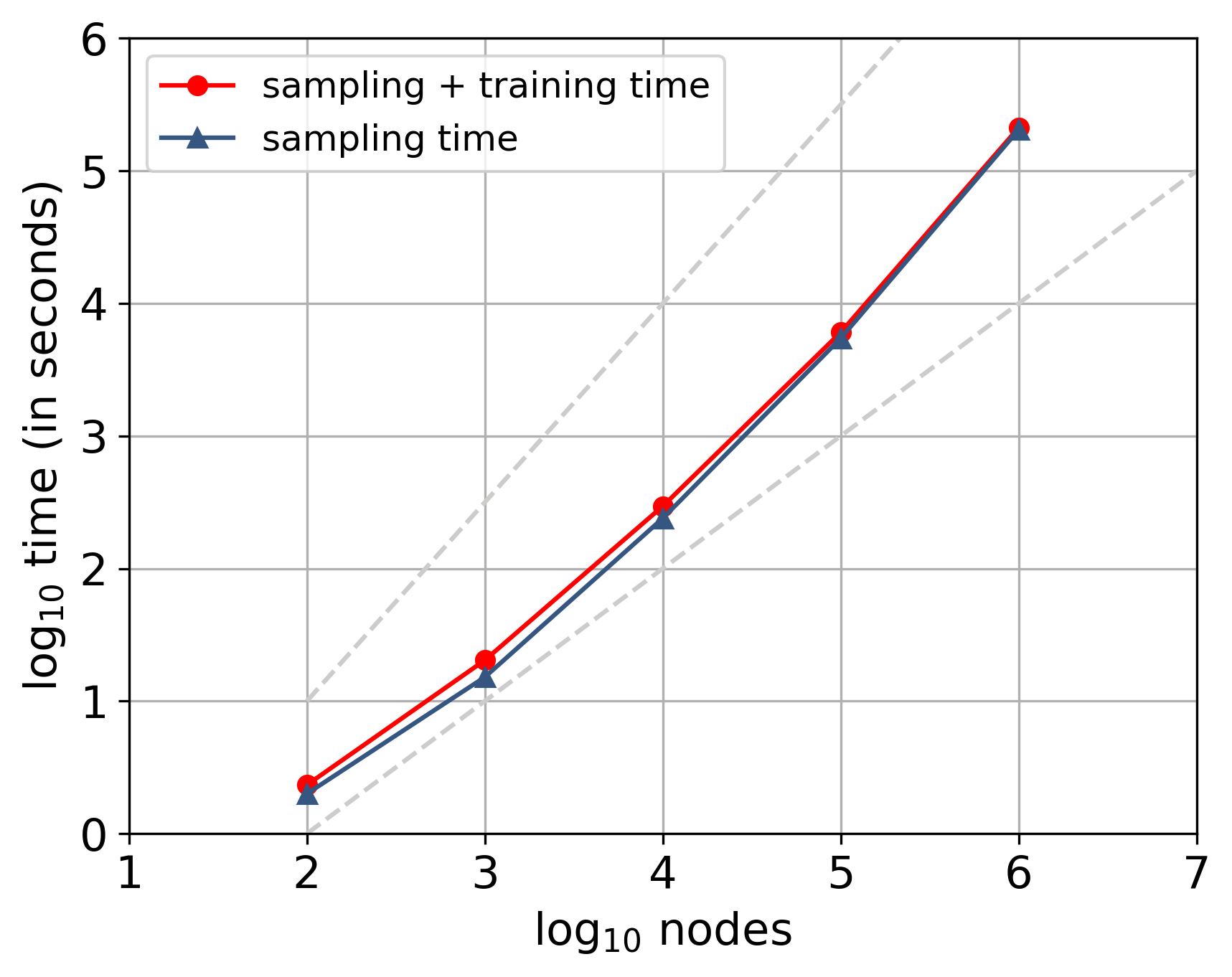} 
\caption{Average execution time of \textit{struc2vec} on Erd\"os-R\'enyi graphs with average degree of 10. Training time refers to the additional time required by Skip-Gram.}
\label{scalability_figure}
\end{figure}

In order to illustrate its scalability, we apply \textit{struc2vec} with the first two optimizations to instances of the Erd\"os-R\'enyi random graph model (using 128 dimensions, 10 walks per node, walk length 80, Skip-Gram window 10). We compute the average execution time for 10 independent runs on graphs with sizes from 100 to 1,000,000 nodes and average degree of 10. In order to speed up training the language model, we use Skip-Gram with Negative Sampling~\cite{word2vecmiko}. Figure~\ref{scalability_figure} shows the execution time (in log-log scale) indicating that \textit{struc2vec} scales super-linearly but closer to linear than to $n^{1.5}$ (dashed lines). Thus, despite its unfavorable worst case time and space complexity, in practice \textit{struc2vec} can be applied to very large networks.

\section{Conclusion}
\label{sec:conclusion}

Structural identity is a concept of symmetry in networks in which nodes are identified based on the network structure. The concept is strongly related to functions or roles played by nodes in the network, an important problem in social and hard sciences. 

We propose \textit{struc2vec}, a novel and flexible framework to learn representations that capture the structural identity of nodes in a network. \textit{struc2vec} assesses the structural similarity of node pairs by considering a hierarchical metric defined by the ordered degree sequence of nodes and uses a weighted multilayer graph to generate context. 

We have shown that \textit{struc2vec} excels in capturing the structural identity of nodes, in comparison to state-of-the-art techniques such as \textit{DeepWalk}, \textit{node2vec} and \textit{RolX}. It overcomes their limitation by focusing explicitly on structural identity. Not surprising, we also show that \textit{struc2vec} is superior in classification tasks where node labels are more dependent on their role or structural identity. Last, different models to generate representations tend to capture different properties, and we argue that structural identity is clearly important when considering possible node representations.



\bibliographystyle{ACM-Reference-Format}
\bibliography{sigproc} 


\begin{thebibliography}{00}


\ifx \showCODEN    \undefined \def \showCODEN     #1{\unskip}     \fi
\ifx \showDOI      \undefined \def \showDOI       #1{#1}\fi
\ifx \showISBNx    \undefined \def \showISBNx     #1{\unskip}     \fi
\ifx \showISBNxiii \undefined \def \showISBNxiii  #1{\unskip}     \fi
\ifx \showISSN     \undefined \def \showISSN      #1{\unskip}     \fi
\ifx \showLCCN     \undefined \def \showLCCN      #1{\unskip}     \fi
\ifx \shownote     \undefined \def \shownote      #1{#1}          \fi
\ifx \showarticletitle \undefined \def \showarticletitle #1{#1}   \fi
\ifx \showURL      \undefined \def \showURL       {\relax}        \fi
\providecommand\bibfield[2]{#2}
\providecommand\bibinfo[2]{#2}
\providecommand\natexlab[1]{#1}
\providecommand\showeprint[2][]{arXiv:#2}

\bibitem[\protect\citeauthoryear{Atias and Sharan}{Atias and Sharan}{2012}]%
        {Atias2012}
\bibfield{author}{\bibinfo{person}{Nir Atias} {and} \bibinfo{person}{Roded
  Sharan}.} \bibinfo{year}{2012}\natexlab{}.
\newblock \showarticletitle{Comparative analysis of protein networks: hard
  problems, practical solutions}.
\newblock \bibinfo{journal}{{\it Commun. ACM}}  \bibinfo{volume}{55}
  (\bibinfo{year}{2012}).
\newblock


\bibitem[\protect\citeauthoryear{Bengio, Ducharme, Vincent, and Jauvin}{Bengio
  et~al\mbox{.}}{2003}]%
        {nnlm}
\bibfield{author}{\bibinfo{person}{Yoshua Bengio}, \bibinfo{person}{Réjean
  Ducharme}, \bibinfo{person}{Pascal Vincent}, {and} \bibinfo{person}{Christian
  Jauvin}.} \bibinfo{year}{2003}\natexlab{}.
\newblock \showarticletitle{A Neural Probabilistic Language Model}.
\newblock \bibinfo{journal}{{\em JMLR\/}} (\bibinfo{year}{2003}).
\newblock


\bibitem[\protect\citeauthoryear{Blondel, Gajardo, Heymans, Senellart, and
  Van~Dooren}{Blondel et~al\mbox{.}}{2004}]%
        {Blondel2004}
\bibfield{author}{\bibinfo{person}{V Blondel}, \bibinfo{person}{A Gajardo},
  \bibinfo{person}{M Heymans}, \bibinfo{person}{P Senellart}, {and}
  \bibinfo{person}{P Van~Dooren}.} \bibinfo{year}{2004}\natexlab{}.
\newblock \showarticletitle{A measure of similarity between graph vertices:
  Applications to synonym extraction and web searching}.
\newblock \bibinfo{journal}{{\em SIAM review\/}} (\bibinfo{year}{2004}).
\newblock


\bibitem[\protect\citeauthoryear{Cao, Lu, and Xu}{Cao et~al\mbox{.}}{2016}]%
        {dnnreps}
\bibfield{author}{\bibinfo{person}{Shaosheng Cao}, \bibinfo{person}{Wei Lu},
  {and} \bibinfo{person}{Qiongkai Xu}.} \bibinfo{year}{2016}\natexlab{}.
\newblock \showarticletitle{Deep Neural Networks for Learning Graph
  Representations}. In \bibinfo{booktitle}{{\em AAAI}}.
\newblock


\bibitem[\protect\citeauthoryear{Fouss, Pirotte, Renders, and Saerens}{Fouss
  et~al\mbox{.}}{2007}]%
        {Fouss2007}
\bibfield{author}{\bibinfo{person}{F Fouss}, \bibinfo{person}{A Pirotte},
  \bibinfo{person}{J Renders}, {and} \bibinfo{person}{M Saerens}.}
  \bibinfo{year}{2007}\natexlab{}.
\newblock \showarticletitle{Random-Walk Computation of Similarities Between
  Nodes of a Graph with Application to Collaborative Recommendation}.
\newblock \bibinfo{journal}{{\em IEEE Trans. on Knowl. and Data Eng.\/}}
  (\bibinfo{year}{2007}).
\newblock


\bibitem[\protect\citeauthoryear{Grover and Leskovec}{Grover and
  Leskovec}{2016}]%
        {node2vec-kdd2016}
\bibfield{author}{\bibinfo{person}{Aditya Grover} {and} \bibinfo{person}{Jure
  Leskovec}.} \bibinfo{year}{2016}\natexlab{}.
\newblock \showarticletitle{node2vec: Scalable Feature Learning for Networks}.
  In \bibinfo{booktitle}{{\em ACM SIGKDD}}.
\newblock


\bibitem[\protect\citeauthoryear{Henderson, Gallagher, Eliassi-Rad, Tong, Basu,
  Akoglu, Koutra, Faloutsos, and Li}{Henderson et~al\mbox{.}}{2012}]%
        {henderson2012rolx}
\bibfield{author}{\bibinfo{person}{K Henderson}, \bibinfo{person}{B Gallagher},
  \bibinfo{person}{T Eliassi-Rad}, \bibinfo{person}{H Tong}, \bibinfo{person}{S
  Basu}, \bibinfo{person}{L Akoglu}, \bibinfo{person}{D Koutra},
  \bibinfo{person}{C Faloutsos}, {and} \bibinfo{person}{L Li}.}
  \bibinfo{year}{2012}\natexlab{}.
\newblock \showarticletitle{Rolx: structural role extraction \& mining in large
  graphs}. In \bibinfo{booktitle}{{\em ACM SIGKDD}}.
\newblock


\bibitem[\protect\citeauthoryear{Kleinberg}{Kleinberg}{1999}]%
        {Kleinberg1999}
\bibfield{author}{\bibinfo{person}{Jon~M Kleinberg}.}
  \bibinfo{year}{1999}\natexlab{}.
\newblock \showarticletitle{Authoritative sources in a hyperlinked
  environment}.
\newblock \bibinfo{journal}{{\em Journal of the ACM (JACM)\/}}
  (\bibinfo{year}{1999}).
\newblock


\bibitem[\protect\citeauthoryear{Leicht, Holme, and Newman}{Leicht
  et~al\mbox{.}}{2006}]%
        {Leicht2006}
\bibfield{author}{\bibinfo{person}{Elizabeth~A Leicht}, \bibinfo{person}{Petter
  Holme}, {and} \bibinfo{person}{Mark~EJ Newman}.}
  \bibinfo{year}{2006}\natexlab{}.
\newblock \showarticletitle{Vertex similarity in networks}.
\newblock \bibinfo{journal}{{\em Physical Review E\/}}  \bibinfo{volume}{73}
  (\bibinfo{year}{2006}).
\newblock


\bibitem[\protect\citeauthoryear{Leskovec and Mcauley}{Leskovec and
  Mcauley}{2012}]%
        {leskovec2012learning}
\bibfield{author}{\bibinfo{person}{Jure Leskovec} {and}
  \bibinfo{person}{Julian~J Mcauley}.} \bibinfo{year}{2012}\natexlab{}.
\newblock \showarticletitle{Learning to discover social circles in ego
  networks}. In \bibinfo{booktitle}{{\em NIPS}}.
\newblock


\bibitem[\protect\citeauthoryear{Lorrain and White}{Lorrain and White}{1971}]%
        {lorrain1971structural}
\bibfield{author}{\bibinfo{person}{Francois Lorrain} {and}
  \bibinfo{person}{Harrison~C White}.} \bibinfo{year}{1971}\natexlab{}.
\newblock \showarticletitle{Structural equivalence of individuals in social
  networks}.
\newblock \bibinfo{journal}{{\em The Journal of mathematical sociology\/}}
  \bibinfo{volume}{1} (\bibinfo{year}{1971}).
\newblock


\bibitem[\protect\citeauthoryear{Mikolov, Chen, Corrado, and Dean}{Mikolov
  et~al\mbox{.}}{2013a}]%
        {skipgram-mikolov}
\bibfield{author}{\bibinfo{person}{Tomas Mikolov}, \bibinfo{person}{Kai Chen},
  \bibinfo{person}{Greg Corrado}, {and} \bibinfo{person}{Jeffrey Dean}.}
  \bibinfo{year}{2013}\natexlab{a}.
\newblock \showarticletitle{Efficient Estimation of Word Representations in
  Vector Space}. In \bibinfo{booktitle}{{\em ICLR Workshop}}.
\newblock


\bibitem[\protect\citeauthoryear{Mikolov, Sutskever, Chen, Corrado, and
  Dean}{Mikolov et~al\mbox{.}}{2013b}]%
        {word2vecmiko}
\bibfield{author}{\bibinfo{person}{T Mikolov}, \bibinfo{person}{I Sutskever},
  \bibinfo{person}{K Chen}, \bibinfo{person}{G Corrado}, {and}
  \bibinfo{person}{J Dean}.} \bibinfo{year}{2013}\natexlab{b}.
\newblock \showarticletitle{Distributed Representations of Words and Phrases
  and their Compositionality}.
\newblock In \bibinfo{booktitle}{{\em NIPS}}.
\newblock


\bibitem[\protect\citeauthoryear{Narayanan, Chandramohan, Chen, Liu, and
  Saminathan}{Narayanan et~al\mbox{.}}{2016}]%
        {subgraph2vec}
\bibfield{author}{\bibinfo{person}{A Narayanan}, \bibinfo{person}{M
  Chandramohan}, \bibinfo{person}{L Chen}, \bibinfo{person}{Y Liu}, {and}
  \bibinfo{person}{S Saminathan}.} \bibinfo{year}{2016}\natexlab{}.
\newblock \showarticletitle{subgraph2vec: Learning Distributed Representations
  of Rooted Sub-graphs from Large Graphs}. In \bibinfo{booktitle}{{\em Workshop
  on Mining and Learning with Graphs}}.
\newblock


\bibitem[\protect\citeauthoryear{Pedarsani and Grossglauser}{Pedarsani and
  Grossglauser}{2011}]%
        {pedarsani2011}
\bibfield{author}{\bibinfo{person}{Pedram Pedarsani} {and}
  \bibinfo{person}{Matthias Grossglauser}.} \bibinfo{year}{2011}\natexlab{}.
\newblock \showarticletitle{On the privacy of anonymized networks}. In
  \bibinfo{booktitle}{{\em ACM SIGKDD}}.
\newblock


\bibitem[\protect\citeauthoryear{Perozzi, Al-Rfou, and Skiena}{Perozzi
  et~al\mbox{.}}{2014}]%
        {Perozzi2014}
\bibfield{author}{\bibinfo{person}{Bryan Perozzi}, \bibinfo{person}{Rami
  Al-Rfou}, {and} \bibinfo{person}{Steven Skiena}.}
  \bibinfo{year}{2014}\natexlab{}.
\newblock \showarticletitle{DeepWalk: Online Learning of Social
  Representations}. In \bibinfo{booktitle}{{\em ACM SIGKDD}}.
\newblock


\bibitem[\protect\citeauthoryear{Pizarro}{Pizarro}{2007}]%
        {Pizarro2007}
\bibfield{author}{\bibinfo{person}{Narciso Pizarro}.}
  \bibinfo{year}{2007}\natexlab{}.
\newblock \showarticletitle{Structural Identity and Equivalence of Individuals
  in Social Networks Beyond Duality}.
\newblock \bibinfo{journal}{{\em International Sociology\/}}
  \bibinfo{volume}{22} (\bibinfo{year}{2007}).
\newblock


\bibitem[\protect\citeauthoryear{Rakthanmanon, Campana, Mueen, Batista,
  Westover, Zhu, Zakaria, and Keogh}{Rakthanmanon et~al\mbox{.}}{2013}]%
        {rakthanmanon2013}
\bibfield{author}{\bibinfo{person}{T Rakthanmanon}, \bibinfo{person}{B
  Campana}, \bibinfo{person}{A Mueen}, \bibinfo{person}{G Batista},
  \bibinfo{person}{B Westover}, \bibinfo{person}{Q Zhu}, \bibinfo{person}{J
  Zakaria}, {and} \bibinfo{person}{E Keogh}.} \bibinfo{year}{2013}\natexlab{}.
\newblock \showarticletitle{Addressing big data time series: Mining trillions
  of time series subsequences under dynamic time warping}.
\newblock \bibinfo{journal}{{\em ACM TKDD\/}} (\bibinfo{year}{2013}).
\newblock


\bibitem[\protect\citeauthoryear{Sailer}{Sailer}{1978}]%
        {Sailer1978}
\bibfield{author}{\bibinfo{person}{Lee~Douglas Sailer}.}
  \bibinfo{year}{1978}\natexlab{}.
\newblock \showarticletitle{Structural equivalence: Meaning and definition,
  computation and application}.
\newblock \bibinfo{journal}{{\em Social Networks\/}} (\bibinfo{year}{1978}).
\newblock


\bibitem[\protect\citeauthoryear{Salvador and Chan}{Salvador and Chan}{2004}]%
        {salvador2004fastdtw}
\bibfield{author}{\bibinfo{person}{S Salvador} {and} \bibinfo{person}{P Chan}.}
  \bibinfo{year}{2004}\natexlab{}.
\newblock \showarticletitle{{FastDTW}: Toward accurate dynamic time warping in
  linear time and space}. In \bibinfo{booktitle}{{\em Workshop on Min. Temp.
  and Seq. Data, ACM SIGKDD}}.
\newblock


\bibitem[\protect\citeauthoryear{Shervashidze, Schweitzer, van Leeuwen,
  Mehlhorn, and Borgwardt}{Shervashidze et~al\mbox{.}}{2011}]%
        {wlkernel}
\bibfield{author}{\bibinfo{person}{N Shervashidze}, \bibinfo{person}{P
  Schweitzer}, \bibinfo{person}{E van Leeuwen}, \bibinfo{person}{K Mehlhorn},
  {and} \bibinfo{person}{K Borgwardt}.} \bibinfo{year}{2011}\natexlab{}.
\newblock \showarticletitle{Weisfeiler-Lehman Graph Kernels}.
\newblock \bibinfo{journal}{{\em JMLR\/}} (\bibinfo{year}{2011}).
\newblock


\bibitem[\protect\citeauthoryear{Singh, Xu, and Berger}{Singh
  et~al\mbox{.}}{2008}]%
        {Singh2008}
\bibfield{author}{\bibinfo{person}{R Singh}, \bibinfo{person}{J Xu}, {and}
  \bibinfo{person}{B Berger}.} \bibinfo{year}{2008}\natexlab{}.
\newblock \showarticletitle{Global alignment of multiple protein interaction
  networks with application to functional orthology detection}.
\newblock \bibinfo{journal}{{\em PNAS\/}} (\bibinfo{year}{2008}).
\newblock


\bibitem[\protect\citeauthoryear{Tang, Qu, Wang, Zhang, Yan, and Mei}{Tang
  et~al\mbox{.}}{2015}]%
        {line}
\bibfield{author}{\bibinfo{person}{Jian Tang}, \bibinfo{person}{Meng Qu},
  \bibinfo{person}{Mingzhe Wang}, \bibinfo{person}{Ming Zhang},
  \bibinfo{person}{Jun Yan}, {and} \bibinfo{person}{Qiaozhu Mei}.}
  \bibinfo{year}{2015}\natexlab{}.
\newblock \showarticletitle{LINE: Large-scale Information Network Embedding}.
  In \bibinfo{booktitle}{{\em WWW}}.
\newblock


\bibitem[\protect\citeauthoryear{Wang, Cui, and Zhu}{Wang
  et~al\mbox{.}}{2016}]%
        {structural}
\bibfield{author}{\bibinfo{person}{Daixin Wang}, \bibinfo{person}{Peng Cui},
  {and} \bibinfo{person}{Wenwu Zhu}.} \bibinfo{year}{2016}\natexlab{}.
\newblock \showarticletitle{Structural Deep Network Embedding}. In
  \bibinfo{booktitle}{{\em ACM SIGKDD}}.
\newblock


\bibitem[\protect\citeauthoryear{Zachary}{Zachary}{1977}]%
        {zachary1977information}
\bibfield{author}{\bibinfo{person}{Wayne~W Zachary}.}
  \bibinfo{year}{1977}\natexlab{}.
\newblock \showarticletitle{An information flow model for conflict and fission
  in small groups}.
\newblock \bibinfo{journal}{{\em Journal of anthropological research\/}}
  (\bibinfo{year}{1977}).
\newblock


\bibitem[\protect\citeauthoryear{Zager and Verghese}{Zager and
  Verghese}{2008}]%
        {Zager2008}
\bibfield{author}{\bibinfo{person}{Laura~A Zager} {and}
  \bibinfo{person}{George~C Verghese}.} \bibinfo{year}{2008}\natexlab{}.
\newblock \showarticletitle{Graph similarity scoring and matching}.
\newblock \bibinfo{journal}{{\em Applied mathematics letters\/}}
  (\bibinfo{year}{2008}).
\newblock


\end{thebibliography}

\end{document}